\documentclass[acmsmall,authorversion,nonacm]{acmart}

\usepackage{color}
\colorlet{RED}{red} % so \todo works in section/subsection headings
\usepackage{booktabs}
\usepackage{comment}
\usepackage{wrapfig}
\usepackage{soul} % for \hl highlight command

\AtBeginDocument{%
  \providecommand\BibTeX{{%
    \normalfont B\kern-0.5em{\scshape i\kern-0.25em b}\kern-0.8em\TeX}}}

\def\fig#1{Figure~\ref{#1}}
\def\tab#1{Table~\ref{#1}}

\def\sec#1{Section~\ref{#1}}

\newcommand{\rev}[1]{#1} % regular text for production version
\newcommand{\revnote}[1]{} % EMPTY nop for production version

\begin{document}

%%
%% The "title" command has an optional parameter,
%% allowing the author to define a "short title" to be used in page headers.
%\title{Groundwork, Orienting, Bridging, Magic, and Coping: \\ The Socioemotional Workflow of Data Scientists Collaborating with Clients in Industry and Academia}

%\title[The Collaborative Workflow of Data Science Consulting Across Industry and Academia]{Groundwork, Orienting, Bridging, Magic, and Coping: \\ The Collaborative Workflow of Data Science Consulting Across Industry and Academia}

%\title[The Collaborative Workflow of Consulting Data Scientists Across Industry and Academia]{Groundwork, Orienting, Bridging, Magic, and Coping: \\ The Collaborative Workflow of Consulting Data Scientists Across Industry and Academia}

%\title{Beyond the Technical: The Outer Loop of Data Science Collaboration Workflows in Industry and Academia}

%\title{Beyond the Technical Inner Loop: The Socioemotional Outer Loop of Data Science Collaboration Workflows in Industry and Academia}

\title[How Data Scientists Navigate the Outer Loop of Client Collaborations in Industry and Academia]{Orienting, Framing, Bridging, Magic, and Counseling: \\ How Data Scientists Navigate the Outer Loop of Client Collaborations in Industry and Academia}

%Orienting, Magic, and Coping: how data scientists work with clients in industry and academia

%Dropping In, Magic, and Coping: data scientists working with clients in industry and academia

%Groundwork, Dropping In, Framing, Magic, and Coping: data scientists working with clients in industry and academia

%\todo{title contender: surfacing the invisible social infrastructure behind data science collaborations across academia and industry}

%%
%% The "author" command and its associated commands are used to define
%% the authors and their affiliations.
%% Of note is the shared affiliation of the first two authors, and the
%% "authornote" and "authornotemark" commands
%% used to denote shared contribution to the research.

%\author{(ANONYMIZED FOR SUBMISSION)}

\author{Sean Kross}
\email{seankross@ucsd.edu}
\orcid{0000-0001-5215-0316}
\affiliation{%
  \institution{UC San Diego}
  \city{La Jolla}
  \state{California}
  \country{USA}
}

\author{Philip J. Guo}
\email{pg@ucsd.edu}
\affiliation{%
  \institution{UC San Diego}
  \city{La Jolla}
  \state{California}
  \country{USA}
}

\begin{abstract}

Data scientists often collaborate with clients to analyze data to meet a client's needs. What does the end-to-end workflow of a data scientist's collaboration with clients look like throughout the lifetime of a project? To investigate this question, we interviewed ten data scientists (5 female, 4 male, 1 non-binary) in diverse roles across industry and academia. We discovered that they work with clients in a six-stage outer-loop workflow, which involves 1) laying groundwork by building trust before a project begins, 2) orienting to the constraints of the client's environment, 3) collaboratively framing the problem, 4) bridging the gap between data science and domain expertise, 5) the inner loop of technical data analysis work, 6) counseling to help clients emotionally cope with analysis results. This novel outer-loop workflow contributes to CSCW by expanding the notion of what collaboration means in data science beyond the widely-known inner-loop technical workflow stages of acquiring, cleaning, analyzing, modeling, and visualizing data. We conclude by discussing the implications of our findings for data science education, parallels to design work, and unmet needs for tool development.

\end{abstract}


%%
%% Keywords. The author(s) should pick words that accurately describe
%% the work being presented. Separate the keywords with commas.

%% A "teaser" image appears between the author and affiliation
%% information and the body of the document, and typically spans the
%% page.
%\begin{teaserfigure}
%  \includegraphics[width=0.6\textwidth]{images/google-images-workflow.png}
%  \caption{xxx}
%  \Description{}
%  \label{fig:teaser}
%\end{teaserfigure}

\maketitle

\section{Introduction}
\label{sec:intro}

%Jupyter is not enough: surfacing the invisible social fabric beneath data science collaborations
%- most prior data science studies, even ones in HCI and collaboration, focus on TOOLS and TOOL MEDIATION (e.g., even Mary Beth Kery's studies, i think? and IBM with Amy Zhang, etc.) ... we focus on the opposite of tools!

%\todo{we have 3 sections before the real jupyter notebook action takes place ... really emphasize that in a diagram or something; joke title: 'everything that happens before the first jupyter notebook launches', or "Jupyter is not enough: understanding the solar system of data science", "Programming/statistics is not enough: blah blah blah", "programming is not enough: the invisible design of data science"}

%\todo{thesis: what kind of effective social infrastructure do data scientist need to build?!? "Jupyter is not enough: how data scientists build sociotechnical infrastructure", "the sociology of doing data science in 2020"}

\begin{figure}
\includegraphics[width=\textwidth]{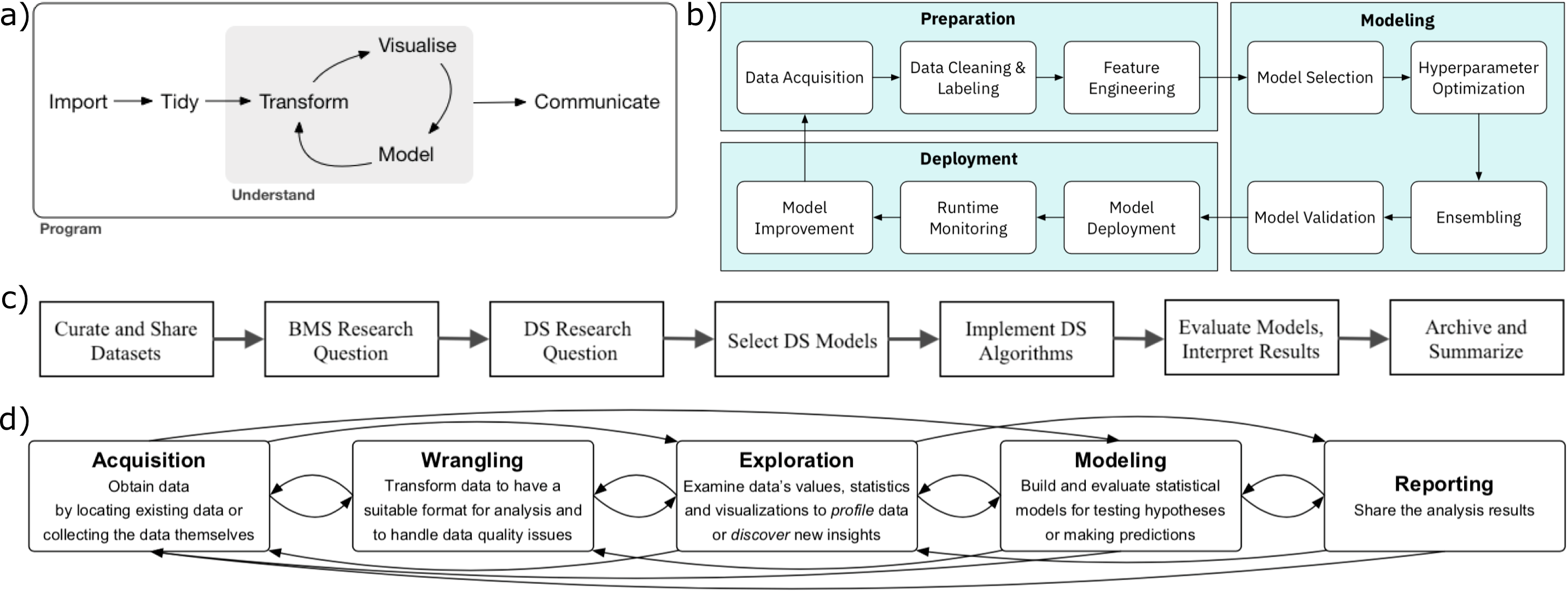}
%\vspace{-1.5em} % stent
\caption{Examples of technical data science workflows from a)~Wickham and Grolemund~\cite{Wickham2017}, b)~Wang et al.~\cite{WangDakuo2019} also used by Zhang et al.~\cite{Zhang2020}, c)~Mao et al.~\cite{Mao2019}, d)~Wongsuphasawat et al.~\cite{Wongsuphasawat2019goals}. In this paper we show how these workflows are actually the inner loop of a more general collaborative data science workflow that we illustrate in \fig{fig:workflow}.}
%\vspace{-0.75em} % stent
\label{fig:krebs}
\end{figure}

%characterizing data science workflows:
%- Guo~\cite{GuoPhD2012}
%- Wickham~\cite{Wickham2017}
%- Dakuo Wang diagram~\cite{WangDakuo2019,Zhang2020}
%- Jeff Heer and friends~\cite{Kandel2012,Wongsuphasawat2019goals}
%- Stoudt~\cite{Stoudt2020principles}
%- Alspaugh~\cite{Alspaugh2019}
%- just do a google image search for "data science workflow" and you'll come up with tons of results

Data science has grown so ubiquitous over the past decade that anyone familiar with the field must have seen workflow diagrams like those in \fig{fig:krebs}, which often appear throughout the literature~\cite{GuoPhD2012,Wickham2017,WangDakuo2019,Wongsuphasawat2019goals,Zhang2020}. A Google image search for `data science workflow' shows many similar diagrams from across the web. While details vary, these  all depict the \emph{technical} workflow of data science, usually with stages such as acquiring, wrangling, cleaning, exploring, modeling, and visualizing data. Many researchers have studied how data scientists work both individually and collaboratively in each of those stages~\cite{Alspaugh2019,Kandel2012,Wongsuphasawat2019goals,Mao2019,Zhang2020,Kim2016} and highlighted the myriad challenges they face; they have also built numerous tools to support all of these stages~\cite{Drosos2020,LauVLHCC2020,Guo2011,GuoPhD2012,Kandel2011,Satyanarayan2014,Wongsuphasawat2016}. % add tool examples for each of these stages

Yet despite all of these advances in our understanding, we argue in this paper that the ubiquitous data science workflows we are familiar with (e.g., \fig{fig:krebs}) \emph{capture only a portion of what data scientists actually do in practice}. Specifically, they represent the tight ``inner loop'' of iterative day-to-day technical work that is required to turn raw data into insights.
But before any technical work can occur, data scientists must find clients to collaborate with and establish the parameters of those working relationships. For instance, in academia a postdoc in statistics may do data science work for a cancer research lab PI on a grant that partially funds their salary. In industry, a freelance data scientist may run their own independent consultancy and work with a variety of external clients; and within large companies or government agencies, data scientists often act as internal consultants to provide analytical services to different clients within their organization.

% original version introduces 'outer loop' a priori
\begin{comment}
Once a working relationship has been established and the project is underway, data scientists periodically interface with their clients in a higher-level ``outer loop'' workflow (e.g., in weekly or monthly client meetings) in addition to the inner loop of their day-to-day technical work as shown in \fig{fig:krebs}.
%
\emph{\textbf{What does this outer loop of data science collaboration with clients look like throughout the lifetime of a project?}}
%
Understanding this outer loop is important for capturing a more complete picture of what data scientists do in practice, since their end goal is often to analyze data to meet a client's needs~\cite{TypeA}. The few prior studies of data science collaborations have focused on the inner loop of how people collaborate within the context of technical workflows~\cite{Kim2016,Mao2019,Zhang2020}, so our study aims to expand this understanding to end-to-end client interactions.
\end{comment}

Once a working relationship has been established and the project is underway, data scientists periodically interface with their clients, usually in weekly or monthly meetings. The few prior studies of data science collaborations~\cite{Kim2016,Mao2019,Zhang2020} have focused on how teams collaborate within the context of technical workflows like those in \fig{fig:krebs}. But since clients are often not involved in the inner loop of day-to-day technical work, their collaborative activities and interpersonal dynamics with data scientists likely take different forms.
\emph{\textbf{What does the end-to-end workflow of a data scientist's collaboration with clients look like throughout the lifetime of a project?}}
Understanding this client-oriented workflow is important for capturing a more complete picture of what data scientists actually do in practice beyond their technical work, since their end goal is often to analyze data to meet clients' needs~\cite{TypeA}.

To investigate this question, we performed semi-structured interviews with ten data scientists (5 female, 4 male, 1 non-binary) in diverse roles across industry and academia. We focused our conversations on how they find clients (or how clients find them), how they navigate these client relationships throughout a project, and what challenges they face in communicating with clients.

\begin{figure}
\includegraphics[width=\textwidth]{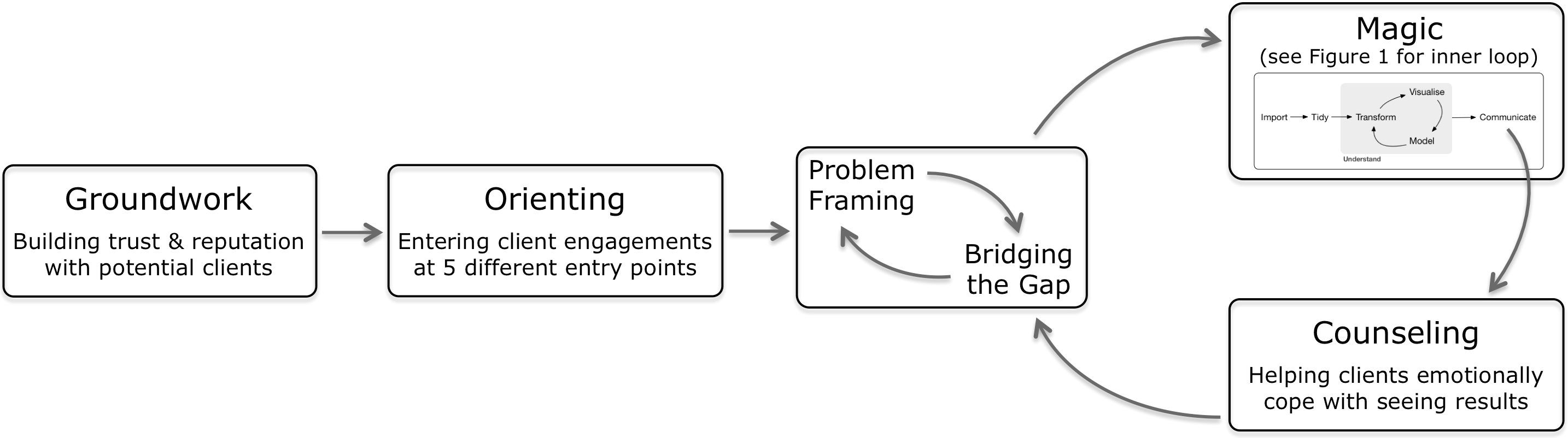}
%\vspace{-1em} % stent
\caption{The outer loop of data science collaboration with clients, which we discovered via an interview study of ten data scientists. The inner-loop technical workflows in \fig{fig:krebs} all reside within the ``magic'' stage.}
%\vspace{-1em} % stent
\label{fig:workflow}
\end{figure}

We distilled our interview findings into an \emph{outer loop workflow} of collaborative data science shown in \fig{fig:workflow}, which subsumes the inner-loop diagrams in \fig{fig:krebs}. Our workflow has 6 stages:

\begin{itemize}
    \item \textbf{Groundwork}: Data scientists must build trust, reputation, and social capital in order to establish working relationships with clients before a project even begins.
    
    \item \textbf{Orienting}: Data scientists join projects at different times and must rapidly orient themselves to the constraints of their given environment. We found five ways that they enter into client engagements: 1)~at the very start of a project, 2)~when the client already has an analysis technique or 3)~data set in mind, 4)~when the client wants them to compute a specific value, and 5)~when the client has already tried and failed to do an analysis themselves.
    
    \item \textbf{Problem Framing}: Data scientists engage in conversations to probe the client's assumptions and work with them to frame the actual underlying problem they want to solve with data.
    
    \item \textbf{Bridging the Gap}: Concurrent with problem framing, data scientists must also bridge the gap between their analytical expertise and the client's domain-specific knowledge.
    
    \item \textbf{Magic}: Data scientists report that clients refer to their day-to-day technical work as ``magic'' since clients often do not see or understand it. The inner-loop technical workflow (e.g., \fig{fig:krebs}) resides within this stage, so that is how our outer-loop workflow subsumes it.

    \item \textbf{Counseling}: When showing results to clients, data scientists must also counsel them to provide emotional reassurance and help them cope with seeing less-than-favorable results.
    
\end{itemize}

Taken together, this outer loop workflow that we discovered in our study (\fig{fig:workflow}) captures the end-to-end collaborative work that professional data scientists engage in from before a client is found all the way to the end of a client engagement. \emph{\textbf{It contributes to CSCW by expanding the notion of what collaboration means in data science beyond the day-to-day technical work}} of \fig{fig:krebs} that has been covered by prior work~\cite{Kim2016,Mao2019,Zhang2020}.

More broadly, our study findings extend the scope of findings from prior CSCW studies in data science collaborations~\cite{Mao2019,Hou2017,Zhang2020,Kim2016}, consultative work with clients~\cite{Stager1986,Lending2013,Tang2019,Gomez2016}, and data management lifecycles~\cite{Ball2012,Miao2017,Pine2015}. Specifically, we contribute novel insights about the extensive groundwork that occurs before collaborations can even begin, the myriad ways that data scientists must orient themselves at the start of projects, and the emotional labor they engage in to manage client relationships in the face of power imbalances (e.g., during the problem framing, bridging the gap, and counseling stages). We discuss how our findings augment prior work in more detail in \sec{sec:related} (Related Work) and \sec{sec:big-one} (Discussion: Relationship to Prior CSCW Work).

We conclude by discussing the implications of our findings for data science education, parallels to design work, and unmet needs for tool development.
First, we discuss how this outer loop workflow leads data scientists to build \emph{socio-emotional infrastructure} to support their clients' needs, which complements the technical infrastructure provided by computer-based tools. However, socio-emotional work such as building trust, orienting into a project's prior constraints, and helping clients emotionally cope with analysis results are not currently taught in most data science curricula, which instead focus mostly on technical education. Next, we draw parallels between the work of data scientists and those of professional designers, who also must work with clients under various constraints. However, unlike designers who have visually-expressive tools like CAD, Photoshop, and Figma, data scientists often lack expressive representations to make their ongoing work transparent and legible to clients; as a result, clients often perceive what data scientists do day-to-day as opaque ``magic,'' which leads to communication barriers. Thus, we identify an unmet need for future data science tools to construct more expressive representations to foster collaborative work.

In sum, this paper's contributions to CSCW are:

\begin{itemize}

\item A novel characterization of client-oriented collaborative work in data science beyond the usual technical workflow of acquiring, cleaning, analyzing, modeling, and visualizing data.

\item Findings from an interview study of ten data scientists who work with a variety of clients across industry and academia, synthesized into a six-stage outer-loop collaborative workflow.

\item A call for more expressive representations to foster collaborative work in data science.

\end{itemize}
%\newpage % stent

\section{Background and Related Work}
\label{sec:related}

Our study was inspired by and extends prior studies of data science collaborations in the HCI and CSCW literature, especially those involving close interactions with clients. It is more broadly related to consultative work in other fields such as design, business, I.T., and statistics.

\subsection{Types of Data Scientists}
\label{sec:dstypes}

% R1: In one case, I hope that the authors will reconsider the literature that they do cite. The concept of Type A and Type B data scientists came from an informal statement in Quora, and has been developed in the blogosphere rather than in the research literature. By contrast, Hayes has considered four roles in data science projects since about 2015 (search for his posts in Business Over Broadway and in CustomerThink. Hayes work is supported by simple statistical analyses of the large-scale Kaggle surveys. Again by contrast, Zhang et al. described five roles based on a medium-scale survey (Proc. CSCW 2020). Kim et al identified a different set of five distinct working styles in data science (Proc. ICSE 2016). To me, the Type A and Type B analysis in this submission is not really necessary, and there are much stronger conceptions available based in real data samples.

% Sean: Talking about type a versus type b data scientists is important. it's easy to dismiss this terminology as "what's going on in the blogosphere" but this is actually how data scientists are talking about themselves. given the importance of the "ecological lens" of work in cscw {citecitecite} it's work bringing thai terminology into the literature alongside how scholars have characterized different data science roles, etc etc

% Me: Cool sounds good! We can keep that part but just augment it with an academic perspective too, which will satisfy reviewer. I may move that into Methods rather than related work.

We first provide context for our study by specifying the types of data scientists that we focus on. We focus specifically on data scientists who work with clients to provide analytical services for them to make data-informed decisions.
In industry these clients may come from other parts of the company (e.g., a data scientist may work with the marketing growth team for a new product line) or from outside (e.g., external business partners). In academia, data scientists are usually in `soft money' positions (e.g., postdoc or research staff) where their salaries are funded by collaborative grants with various PIs (usually faculty); these academic data scientists may need to juggle multiple grant-funded projects to pay their salaries and thus concurrently work with several clients in different subfields.
Client project engagements usually last from a few months to a few years.

This type of client-oriented role is common throughout both industry and academia. For instance, Zhang et al.~\cite{Zhang2020} found five distinct roles on data science teams in industry, including communicator, manager/executive, domain expert, researcher/scientist, and engineer/analyst/programmer. Under this taxonomy, the data scientists we study are `analysts' whose clients are managers/executives and domain experts. 
Within academia, Mao et al.~\cite{Mao2019} studied the collaboration between data scientists and their clients who were biomedical researchers.
Kim et al.~\cite{Kim2016} further honed in on five types of working styles of industry data scientists: insight providers, modeling specialists, platform builders, polymaths, and team leaders. The data scientists we study are mostly insight providers who \emph{``with a strong background in statistics [their] main task is to generate insights and to support and guide their managers in decision making.''}~\cite{Kim2016} Similarly, using a broader survey of 19,000 data scientists on Kaggle, Hayes found that the most common role was to \emph{``analyze and understand data to influence product or business decisions''} (63\% of respondents)~\cite{Hayes2020}. Some industry practitioners also use the term `Type-A'~\cite{TypesAB,TypesAB2} data scientist to denote this type of \emph{Analyst} role.

In contrast, other types of data scientists (whom we did not study for this paper) function more like software engineers who build data-driven platforms and apply machine learning to create predictive models. Some still work with clients, but others have more fixed long-term roles to build specific software infrastructure, such as Netflix's movie recommendation engine. In Kim et al.'s taxonomy~\cite{Kim2016}, they are platform builders and modeling specialists. In Zhang et al.'s taxonomy~\cite{Zhang2020}, these are researchers/scientists (usually with a machine learning background). Hayes found that 35\% of Kaggle survey respondents built machine learning infrastructure~\cite{Hayes2020}. And some practitioners use the term `Type-B'~\cite{TypesAB,TypesAB2} data scientist to denote this type of \emph{Builder} role.

%The term `data scientist' elicits very different images for different people. Some might think of an applied statistician who uses R or Stata to make statistical inferences; but others might think of a machine learning Ph.D. who writes high-performance C++ code to process terabyte-scale data to build predictive models.

%\subsection{Studies of Data Science Collaborations}
\subsection{\rev{Inner-Loop Collaborative Data Science Workflows}}
%\subsection{\rev{Inner-Loop Studies of Data Science Teams}}
% Or:
%\subsection{\rev{Software-Focused Studies of Data Science Teams}}

The bulk of research around data science centers on the technical workflow shown in \fig{fig:krebs}. Much of this research involves tools to facilitate all of these workflow stages, such as data wrangling systems~\cite{Kandel2011,Guo2011}, computational notebooks~\cite{LauVLHCC2020}, coding assistance tools~\cite{Drosos2020}, and data visualization interfaces~\cite{Wongsuphasawat2016,Satyanarayan2014}. Our study does not explicitly focus on tool use, although we discuss implications for collaborative tool design in \sec{subsec:represent}. Rather, we studied how data scientists and their clients collaborate, so the closest related work to ours are those that study data science collaborations.

%\rev{\sec{subsec:represent} discusses the implications of our findings on collaborative tool design.}

% R2: Focus on tools: While the authors emphasize throughout the paper that this study does not focus on tools, the contribution and discussion is around the lack of tool support in the data science workflow. This seems contradictory.

In industry, two notable studies of data science collaboration were both done at large multinational technology companies: Zhang et al.\ conducted a survey of 183 IBM employees who worked in data science related roles (whom they call \emph{data science workers}~\cite{Muller2019} to emphasize that people with varying job titles work closely with data) to investigate collaboration practices~\cite{Zhang2020}. Kim et al.\ interviewed 16 data scientists at Microsoft about how they collaborate with software engineering teams in particular~\cite{Kim2016}. In academia, Mao et al.\ interviewed 22 collaboration participants containing a mix of data scientists and biomedical domain scientists working together on research projects.

These studies all took a holistic approach of characterizing many facets of collaboration, including who usually collaborated with whom, what roles each collaborator played, how multi-domain teams were structured, and what tools they used to work together. They provide important context for the types of data scientists that we study; as we discussed in \sec{sec:dstypes}, we focus on the `analyst' persona from Zhang et al.~\cite{Zhang2020} and the `insight provider' persona from Kim et al.~\cite{Kim2016}.

Most notably, these studies focused on how collaboration took place within the inner-loop technical workflow stages of \fig{fig:krebs}: Kim et al.'s study listed stages such as collecting data, cleaning data, building predictive models, hypothesis testing, and operationalizing models~\cite{Kim2016}. Zhang et al.'s study referenced the workflow in \fig{fig:krebs}b and emphasized collaboration within this workflow in its paper abstract: \emph{``We found that data science teams are extremely collaborative and work with a variety of stakeholders and tools during the six common steps of a data science workflow (e.g., clean data and train model).''}~\cite{Zhang2020}
Mao et al.~\cite{Mao2019} presented a similar inner-loop technical workflow diagram for academic collaborations between biomedical scientists and data scientists, shown in this paper as \fig{fig:krebs}c.

Our study complements and extends this lineage of work by zooming out of the usual technical workflow stages to consider how data scientists interface with their clients at a higher level both before and during an analysis.
While prior studies characterized the day-to-day technical aspects of collaborative work (e.g., who takes on what roles, what tools they use), our study focuses on the \emph{strategic and emotional considerations} of how data scientists collaborate with clients throughout the lifetime of a project.
\textbf{\emph{Our novel contribution is to synthesize a more general workflow showing an outer loop of analyst-client interactions}} (\fig{fig:workflow}) that progresses from laying the groundwork before an analysis even begins to orienting to problem framing to doing ``magic'' (which is what many of our participants' clients referred to their technical work as) to helping clients cope with sub-optimal analysis results. The inner-loop technical workflows that prior studies were all situated within appear inside the ``magic'' node of our more general workflow diagram.

\subsection{How Data Scientists Work with Clients}

The closest related projects to our study are those that not only study general data science collaborations (see prior section) but that also zoom in on how data scientists work with clients.

Mao et al. interviewed 16 data scientists and 6 of their clients who were biomedical scientists in academia~\cite{Mao2019}. They analyzed client interactions through the theoretical lens of the Olsons' framework from \emph{Distance Matters}~\cite{Olson2000} and found a variety of challenges in building common ground in these collaborations. Our study corroborates some of their findings, especially as expressed in the Problem Framing and Bridging the Gap stages of our outer-loop workflow (\sec{sec:frame}). Specifically, data scientists and their clients face a tension when defining the actual problem to solve and need to balance both knowledge from general data-analytic methods and from the target domain. Mao et al. elegantly expressed this tension as one between `find the right answer' versus `ask the right question.' We add to their contributions here by making a connection from data science to what professional designers do in the design thinking process~\cite{Cross2011,Cross2018} when iteratively framing the problem. In \sec{sec:big-one} of the Discussion, we further contextualize our findings in relation to this work.

Our study extends the scope of Mao et al.'s work in several main ways. First, we characterize the legwork that data scientist must do \emph{before a collaboration even begins} (the Groundwork stage in \sec{sec:groundwork}) as well as the five main ways in which they enter into collaborations (the Orienting stage in \sec{sec:orienting}). In contrast, Mao et al.'s study focuses mostly on collaborations that are already underway. That said, 
they do discuss aspects of collaboration readiness such as motivations for collaborating on data-centric open science projects in academia. We extend those findings by detailing how data scientists build reputation and trust in themselves and then use five distinct entry points (\sec{sec:orienting}) to enter into collaborations throughout large companies, startups, and academic institutions. Next, we surfaced the variety of emotional labor~\cite{Grandey2000} involved in data scientists doing problem framing (\sec{sec:prob-framing}) and counseling clients (\sec{sec:counseling}) when results look unfavorable, especially when there are power imbalances.
Furthermore, our study covers different but complementary topics as Mao et al. in terms of the `inner loop' when a collaboration is well underway: Mao et al.\ use the Olsons' framework to characterize technology readiness of how specific software tools can mediate analyst-client interactions. Our study does not focus on tools but instead covers new topics such as how data scientists handle the emotional reactions of clients to their technical `magic' work in the Magic and Counseling stages (Sections~\ref{sec:magic} and~\ref{sec:counseling}, respectively). In sum, our study augments Mao et al.'s findings by synthesizing an end-to-end workflow of how data scientists lay groundwork, enter into collaborations, and emotionally interface with clients.

Another related study was done by Hou and Wang, who performed observations and interviews at two civic data hackathons that connected data scientists with clients from nonprofit organizations (NPOs)~\cite{Hou2017}. They found that having a group of student volunteers serve as ``brokers'' (i.e., intermediaries) between data scientists and NPO representatives helped the two groups work better together, especially during the short timespans of hackathons that last for one or two days. Our setting differs in that data scientists collaborate with clients over longer time frames (a few months or years) in a variety of industry and academic settings, rather than in a short hackathon with NPOs. Also, it is up to the data scientist themselves to ``broker'' those client interactions without third-party involvement; we depict this brokering activity as the Bridging the Gap stage in \fig{fig:workflow} (\sec{sec:bridging}). Lastly, we cover how data scientists enter into such collaborations in the groundwork and orienting stages, whereas Hou and Wang focus on what occurs when stakeholders (e.g., NPOs, client teams, and data scientists) have already been established for the hackathon event.

Related to above, more distantly-related CSCW studies of data scientists working with open government data for social good~\cite{Choi2017} and crisis-response scenarios~\cite{Hellmann2016} also reveal the importance of broker roles in establishing common ground between collaborators. However, those study settings are more decentralized, often rely on part-time and volunteer workers, and involve a vast array of stakeholders, in contrast to the tightly-focused relationship between a data scientist and their client that we study in this paper. Similarly, CSCW studies of large-scale, multidisciplinary, geographically-distributed scientific collaborations (sometimes called e-Science, e-Research, or collaboratories) reveal the socioemotional challenges of remotely coordinating domain scientists and software developers in high-performance computing~\cite{Jirotka2013,Lawrence2006,Olson2008}. Our study differs in that it focuses on finer-grained direct interactions between a data scientist and their client in a colocated setting. In \sec{sec:big-one} we further discuss our findings in relation to these lines of work.

\subsection{Other Types of Consultative Work with Clients}

Zooming out beyond data science, our study continues the lineage of CSCW research on consultative work involving clients in other fields. The closest analogue is design consulting, where UX or product designers are brought in to help ideate and prototype with various business stakeholders~\cite{Cross2011,Cross2018,Gomez2016}. Similar to our data scientists, design consultants must lay groundwork to find clients (\sec{sec:groundwork}) and engage in extensive problem framing (\sec{sec:prob-framing}) before even starting to prototype. However, data science work has a larger variety of entry points (\sec{sec:orienting}) and involves techniques that are more opaque (\sec{sec:magic}) than design work, whose processes and outputs are usually visible to clients. Our setting also shares some similarities with I.T. (information technology) consulting, such as Stager's study at a university computing center~\cite{Stager1986} and Lending and Dillon's work in industry~\cite{Lending2013}; in particular, I.T. consulting also involves groundwork and problem framing, as \emph{``users often do not know what type of information or services a consultant is able to provide''}~\cite{Stager1986}. More distantly-related are business and management consulting, which often involve training consultants to empathize with clients~\cite{Tang2019} and using workflow tools to coordinate business metrics and deliverables~\cite{Mazzoleni2009}; data science work shares some similarities but also involves more perceived technical `magic' (\sec{sec:magic}) than business consulting, which is more easily understandable by clients who are business stakeholders. Finally, since some branches of data science come from academic statistics departments~\cite{Donoho2017}, statistics consultants within universities are similar to modern-day data scientists in some ways since they help researchers to design experiments, apply the appropriate statistical tests, and interpret results~\cite{Bross1974,Finney1982,Lurie1958}; however, statisticians usually work with more fixed data sets and engage in less data wrangling and programming work than data scientists do. Overall our study differentiates itself from prior work in other fields by revealing domain-specific insights about how data science consultants work with clients and synthesizing them into an end-to-end workflow.

\subsection{Data Management Lifecycles}

In a separate lineage of work, researchers in data management characterized data-oriented workflows from a \emph{data-centric} perspective rather than from a \emph{people-centric} perspective as we have done. For instance, Ball summarized half a dozen data management lifecycles (i.e., workflows)~\cite{Ball2012} and Miao et al.\ extended this to data lifecycles for machine learning~\cite{Miao2017}. These lifecycles trace data from initial creation through to various stages of processing, archiving, backup, and distribution.

Even though our study focuses on the interpersonal interactions between data scientists and clients (i.e., a people-centric perspective), it intersects with these data-centric lifecycles at several points. During the Orienting stage (\sec{sec:orienting}) data scientists must enter into a project and quickly figure out whether data has already been collected and if they can influence the collection process. This corresponds to data acquisition, appraisal, and selection phases of some lifecycle diagrams~\cite{Ball2012,Miao2017}. Nearly all other data management lifecycle activities (e.g., versioning, cleaning, archiving, backup) occur within the Magic stage of our workflow since clients are not closely-involved in managing data itself; they are more interested in seeing actionable results from data. The notable exception is clients making decisions in terms of what to measure and which variables are the most important for their needs, which occurs in the Problem Framing stage. As Pine and Liboiron point out, the act of selecting what to measure is infused with politics~\cite{Pine2015}; in our case, it potentially reflects the relative power of clients over the data scientists they collaborate with.

In sum, our study findings augment data management lifecycles with a novel \emph{people-centric} perspective by illustrating the interpersonal dynamics surrounding data. Our workflow covers the socio-emotional challenges of data science work that occur in stages such as Groundwork, Orienting, Problem Framing, and Counseling, which are \emph{not} captured by data management lifecycles.
\section{Methods}
\label{sec:methods}

For this study we interviewed 10 data scientists across industry and academia who regularly collaborate with a variety of clients to provide analytical services for them (see \sec{sec:dstypes} for details). We recruited participants online via social media posts and  snowball sampling from our professional networks. We strove for a diverse sample in terms of demographics, work environments, and client profiles.

Each interview was conducted by the lead author remotely on Zoom and lasted from 1 to 1.5 hours. Participants were not paid. Interviews were semi-structured and focused on how participants interacted with their clients throughout the lifetime of data science projects. The interviewer encouraged, but did not require, participants to show us artifacts from past projects (e.g., business presentations, academic publications) to help ground the conversations. Guiding questions included:

\begin{itemize}
    \item At your workplace, how do new data science projects come to you? Do you proactively look for clients, or do clients find you?
    \item How do you establish new client relationships?
    \item Walk me through the course of a typical client engagement.
    \item What, if any, difficulties in communication have you had with clients in the past?
    \item Have you ever had to deliver bad news to a client? If so, describe how that went.
%    \item What does `success' mean when working with a client?
\end{itemize}

Most notably, we kept conversations at the level of their client interactions rather than diving into inner-loop technical mechanics of how they use specific data science tools (e.g., \fig{fig:krebs}).

\subsection{Interview Participant Backgrounds}

\begin{table}

\begin{tabular}{@{}lllll@{}}
ID & G & Sector & Workplace & Typical Clients \\
\midrule

P1 & F & Industry & Healthcare startup & business executives, software engineers \\
P2 & F & Academia & Dept.\ of biostatistics & medical doctors, biomedical research PIs \\
P3 & F & Academia & Dept.\ of medicine & medical doctors who have research grants \\
P4 & M & Academia & Medical research center & faculty PIs in biology and statistics \\
P5 & F & Industry & Well-known tech company & junior and senior executives, manager \\
P6 & NB& Industry & Freelance data consultant & companies and government agencies \\
P7 & M & Academia & Dept.\ of biostatistics & PIs in biology, genetics, and statistics \\
P8 & F & Industry & Federal government agency & government representatives, managers \\
P9 & M & Industry & Data-driven health startup & junior and senior executives, manager \\
P10& M & Academia & Brain research center & faculty PIs in various fields \\
\bottomrule

\end{tabular}

\caption{The 10 data scientists in our study, their demographics, workplace, and who their clients typically are. G=gender (F=female, M=male, NB=non-binary). PI=Principal Investigator on a research grant.}

%\vspace{-1em} % stent

\label{tbl:participants}
\end{table}

\tab{tbl:participants} shows our 10 participants' diverse demographic and professional backgrounds. 5 are female, 4 male, and 1 non-binary; and they are evenly split between industry and academia. Within industry, workplaces include startup companies, a well-known large tech company, a government agency on environment protection, and an independent consultant (P6). Participants in academia usually work at large research universities or medical research facilities.
Our participants had between 5 to 25 years of firsthand experience doing data science work\footnote{includes time spent working in various data analysis and statistics roles before `data scientist' became an official job title}.

P1, P5, P8, and P10 work on a team of data scientists, while the other six work individually. We did not investigate teamwork dynamics in this study, though, since we focused on client interactions.
The `Typical Client' column in \tab{tbl:participants} shows where clients usually came from. In industry, clients include various managers, business executives at different levels, and sometimes software engineers. In academia, clients are mostly PIs (Principal Investigators) who hire our participants to do data analysis and statistics consulting for a particular grant-funded project.

%; our participants mostly worked with PIs in the life sciences and medical research fields.

%
%These clients may be their direct supervisors (e.g., principal investigators in academia or managers in industry) [P1, P4, P5, P6, P8, P9, P10], other members of the organization they work in who are not their supervisors (e.g., research staff in academia or software engineers in industry) [P1, P4, P5, P8], and external clients who are not a part of their organization (e.g.,  small businesses, entrepreneurs, independent consultants) [P1, P2, P3, P4, P6, P7, P9, P10].

\subsection{Data Overview and Analysis}

The lead author recorded notes and prominent quotes during each interview. In addition, all interviews were recorded, transcribed, and viewed by the entire research team (two members). As we studied the videos and transcripts, we iteratively categorized our observations together into major themes using an inductive analysis approach to build a grounded theory~\cite{Corbin2008} of how these data scientists collaborated with their clients. Two researchers performed open coding on interview transcript data as it arrived and met regularly to reconcile our codes when there were disagreements. We decided to stop after 10 participants since we felt we had reached a reasonable saturation point~\cite{Saunders2018}: several participants often brought up the same themes without any prompting by the interviewer, and it became more difficult to find distinctly new themes.
\rev{
Fifteen intermediate themes emerged during this iterative process, which we eventually merged into six stages. Example intermediate themes included participants' discussions about managing client expectations, defining success in a collaboration, and details about how new projects are first presented to them.
}
After all interviews were completed, we collaboratively merged our codes into a hierarchy that resulted in six final high-level themes (Groundwork, Orienting, Problem Framing, Bridging the Gap, Magic, and Counseling), which we used to construct the workflow diagram shown in \fig{fig:workflow}.
% PG: i tried to play with IRR wording but didn't find a good way to fit it in, so i commented it out for now
%\rev{
%Both research team members reviewed the interview transcripts together during %synchronous meetings, so we did not collect inter-rater reliability %scores~\cite{Nora-IRR}.
%}
%
\rev{
During our meetings, we initially discussed having more stages in our model such as a `success/failure' node at the end. But we eventually decided that negotiating the meaning of success happens throughout different stages (most notably, Counseling) so we interspersed those findings into our six stages.
}

% R1: I found the qualitative analysis to be convincing. There was evidence to support each major point, and the evidence came from more than one informant in almost all cases - thus supporting the authors' claims of saturation. In these terms, I thought that the small sample size was not a problem, precisely because of the paper's discussion of saturation.

% R1: This is a NIT. The paper claims an inductive method, and cites Corbin and Strauss. Other aspects of the reasonably well-documented qualitative method seem very much aligned to grounded theory, which is the principal concern of Corbin and Strauss. Was there a reason why the analysis in this paper was not described as grounded theory?

% R2: 2. Interview analysis. While the main contribution of the paper is a qualitative analysis of the interviews, the details of the analysis appear to be unclear from the writing. More details have to be provided, such as the number of coders, and a description of each iteration in ``After several rounds of iteration and merging’’.

% - Beth Simon has a paper about grounded theory for CS Education
% - see p191 in 'qualitative methods' chapter of Computing Education Research handbook has more details on inductive categorization

\subsection{Study Design Limitations}

Although we strove to include participants from a variety of demographic and professional backgrounds, our personal recruitment and snowball sampling led to some limitations: Everyone was from North America, those in academia usually worked with PIs in the life sciences and medical research, and those in industry may have not been able to disclose all the details of their work due to corporate confidentiality agreements since we did not work at the same companies that they did.

We concluded our study after interviewing 10 participants since we felt we had reached a reasonable saturation point~\cite{Saunders2018} (e.g., several participants independently brought up the same themes, it became harder to find new themes). Since this was a qualitative study, we were not trying to quantify event occurrences or divide our interview data into segments. However, we acknowledge that having more participants in more varied settings may have yielded additional findings. In addition, we believe our study population is representative of the common types of client-facing analysts roles described in \sec{sec:dstypes}. However we did not cover machine learning or platform-building oriented roles, so our findings may not generalize to those types of data scientists.

Furthermore, we focused on how data scientists interact with their clients, but we did not account for greater structural issues such as how their work environments, team dynamics, or organizational hierarchies might have affected their workflows. Related work by Zhang et al.~\cite{Zhang2020}, Kim et al.~\cite{Kim2016}, and Mao et al.~\cite{Mao2019} cover these organizational aspects of team-based interactions.

Finally, we did not interview the clients whom our participants worked with, so we can report only from the perspectives of data scientists. Thus, our findings are representative of what data scientists think their clients' needs, problems, and concerns are throughout their engagements.

\section{Groundwork: Building Trust with Clients in Positions of Power}
\label{sec:groundwork}

We report findings in the order we illustrated in \fig{fig:workflow}. First, interview participants emphasized the importance of building trust with potential clients before a collaboration can even begin.

\subsection{Finding clients and building initial trust}

Clients seek out data scientists based on 
trust built from prior working relationships. For example, P1 mentions how clients typically find her:

\begin{quote}
\textit{I've noticed the people that I have a closer relationship with, maybe we've worked together on different projects in the past, they feel really comfortable asking me directly for things or they feel like they know the path to me.}
\end{quote}

Some need to proactively seek out clients, though. For instance, P8 works on a data science team in a governmental agency on environmental protection and describes the publicity work she must do internally to advertise their services:

\begin{quote}
\textit{Now we are sort of, let's say marketing our team to some degree. I think a lot of that comes through things like internal blog posts, community-of-practice websites, obviously presentations, but you only have so much time where you can give webinars or go to another ministry or go to another group.}
\end{quote}

Once a potential client is identified, data scientists must build trust by showing that they care about the client's domain-specific goals. 
P1 worked for over a decade at a research hospital and recently joined a healthcare startup; she mentioned how it is important in both academia and industry to demonstrate her motivation to understand her client's domain:

\begin{quote}
\textit{That helps you show I am interested in this business and I know how to bring value to this business. I'm not just like churning out numbers, whether that's churning out a p-value for your academic publications or churning out a dashboard -- yet another dashboard -- for your industry stuff.}
\end{quote}

Clients also have their own ``clients'' who come from a broader audience that consumes the final results of an analysis.
For instance, in academia the client may be a lab PI, but the audience for analysis results includes readers of the publications produced by the lab; and in industry the client may be a division director, but the audience is fellow business executives who attend presentations that the director gives.
The end-user audience who is consuming analysis results will also likely have similar domain knowledge as the direct client.
To that end, P2 mentioned empathizing with the audience as well:

\begin{quote}
\textit{If you want to be successful, like if you want to be good at your job, essentially you have to understand that there's an audience who's consuming that analysis. And your goal is to get them to believe you, to trust you, to have confidence that what you're producing makes sense.}
\end{quote}

\subsection{Power dynamics in establishing client relationships}

Clients are often in positions of power (e.g., a senior PI in academia or a business executive in industry), so data scientists must also take power dynamics into consideration when establishing these relationships. For instance, some clients may implicitly expect a foregone conclusion, such as 
a senior design manager who believes one variant of a website's homepage will lead to higher sales volume compared to another variant of the same page, and they want to get some data analysis results to back up their hunch.
%
%an executive wanting to present evidence that their product line is growing at a profitable level that exceeds expectations
Thus, data scientists must balance the client's expectations with their sense of duty to be faithful to the rigor of the analysis process.
Similarly, P4 mentioned that it is important to be able to give honest feedback to their client to help them avoid unrealistic expectations before entering into a collaboration, but power dynamics can make it hard to do so:

\begin{quote}
\textit{I think you should always respect your collaborators, but you need to also understand at times they want you to be their check. They want you to push back, but it's very difficult when there's a huge power difference.}
\end{quote}

The way to build this trust is to have a proven track record of doing high-quality work and accruing social capital within the organization or broader data science community. As P5 explains,

\begin{quote}
\textit{I think that it was really important to have built up credibility, like an institutional reputation, so people believed there were a lot of things I was right about before, or I had given an answer that people believed and made sense to them. Then to come to this kind of situation and to be able to say, um, ``I'm sorry, we can't do this.'' Then I have to spend the capital. It's like you earn the capital and then you spend it.}
\end{quote}

Seven participants (P1, P3, P4, P5, P8, P9, P10) mentioned similar experiences in terms of earning and spending social capital to manage client expectations.

%in the face of power imbalances.

%P1 explains how she must navigate this situation: \emph{``they have to trust you enough to know that you're asking questions because you want to understand the right problem and make sure that the solution is correct or that you have the best interest of them or the business or whatever the context is, at heart.''}

%\begin{quote}
%\textit{It takes a level of trust because they kind of think, either consciously or subconsciously, they've solved the problem. They just can't punch all the right buttons to get the solution, but they have to trust you enough to know that you're asking questions because you want to understand the right problem and make sure that the solution is correct or that you have the best interest of them or the business or whatever the context is, at heart.}
%\end{quote}

\section{Orienting: Five Entry Points into Data Science Collaborations}
\label{sec:orienting}

%\todo{SK: 'dropping in' feels a bit too informal and flippant. maybe just go with 'parachuting in'. ok let's go with 'orienting' for now}

%\todo{if we go with ORIENTING, then when Sean passes thru this section, frame the conversations in terms of orienting and getting their bearings and wayfinding, etc.}

Having established trust (see previous section), a data scientist joins an analysis project at different stages depending on the client's needs. One common sentiment amongst our participants was that they felt they were ``parachuting in'' to a foreign environment whose constraints had already been established before they joined, so they needed to \emph{orient themselves} before starting the technical analysis work. We found five entry points into data science collaborations, each with different kinds of communication that occur between data scientists and clients during this orienting process.

\subsection{At the very start of a project}

% SKTODO: set the stage for P1, look at how one person can have these very difference experiences!
Some data scientists enter an analysis project before anything has been decided or any data has been collected, and they are often relied upon throughout the entire lifetime of the project. This is the ideal case in many of our participants' minds since they get to engage clients in substantive conversations about how to set up the data collection, experiment, or analysis from the ground up.
However, the reality is that oftentimes the client has already established some constraints before bringing them in.
For instance, P1 has over ten years of experience playing different data science roles in medical research and healthcare settings; here she describes the ideal case of being brought in early, along with a range of other entry points that we will describe later in this section:

\begin{quote}
\textit{When [my client] would ask me for things, it would range from ``we need you to design this study'’ to ``we need you to help define the research question’' to ``we need you to make this figure because we want to put this figure in a journal [paper]'' or, ``Hey, can you do this re-analysis because reviewer two wanted this re-analysis?’'}
\end{quote}

% PG: not sure how relevant this below quote is ...
%The client's desire for this wide range of data products \todo{CITE} (a study design, a set research questions, a figure, a summary statistic) is reflected in the wide range of levels of preparedness and understanding that they have in terms of what kinds of work and data will be necessary to make those data products. As P1 continues to explain, a client may tell her:
%
% SKTODO: set up this quote
%
%\textit{```Hey, I want to work closely with you. I understand we have to get very crisp on this definition so that everybody agrees what we're measuring and we need to collect this data, like all these data points, because this is exactly what we need.'}
%
% SKTODO: insert connective tissue
%

\subsection{Client already has an analysis technique in mind}

Some clients bring in a data scientist at the start of a project, but they already have a technique in mind that they want to try out. For instance, a biomedical lab PI may have heard about the latest advances in neural network-based deep learning for image recognition in medical applications, and they want to apply deep learning to detecting a certain kind of cancer cell that they study in their lab. 
% This is a perfect example actually
Their hope is that these new techniques could reveal hidden insights that could lead to, say, notable publications in academia or product growth in industry.
The main upfront conversation that a data scientist must have with clients in this case is to help them manage expectations about their desired technique. For instance, P3 said this about her clients in a medical research context:

%, who uses data science to develop public health interventions as a medical school research faculty, said that:

\begin{quote}
\textit{I feel like people have over-hyped data science, artificial intelligence, these words, I hate even using them now. Because people expect very glamorous sorts of results. So now since I do a lot of question-asking upfront, I see that people have an expectation from all the hype, and then when they don't get it they feel a little let down, and they're disappointed.}
\end{quote}

Similar to the power dynamics mentioned in the prior section, it can be hard for data scientists to push back against a client's expectations when the client is in a position of greater authority and often funding the work.
Thus, our interview participants sometimes offer to start with applying simpler methods (e.g., general linear regression models, gradient boosted decision trees) before agreeing to invest the large upfront time in applying more complex methods such as deep learning.

\subsection{Client already has a data set in mind}

Some clients bring in a data scientist after they have already collected data. In academia this data may come from a variety of lab experiments or from data sets obtained from external collaborators. And in industry this data often comes from usage logs of how customers are using their product (e.g., server logs for web applications, telemetry for mobile apps or internet-connected devices).

These clients have already put in lots of upfront effort and cost to collect the data, so they want to somehow ``capitalize'' on it. P9 recounted such a situation where a client had collected thousands of log entries from wearable fitness devices which their participants had worn for several years:

%; the client wanted to analyze the logs for their research in human metabolism:

\begin{quote}
\textit{I recall one instance of someone saying like, ``I have all this data and I want to show that it's significant.'' And I had to have moments of being like, well, they're very ``significant'' because you spent all this time running experiments, collecting that, they're very meaningful, but let's maybe unpack what that means.}
\end{quote}
What P9 meant by `significant' here is that he is reassuring the client that there is probably \emph{something} meaningful that can come out of this data, although he cannot guarantee any specific set of results.
There is a delicate balance between giving the client emotional reassurance that their efforts in collecting that data has been worthwhile and also setting realistic expectations that real-world data is messy and incomplete, so they cannot guarantee any  desired outcomes. Again, in the ideal case the data scientist would have been involved in the project from the very beginning to help design the data collection process itself, but they are often parachuted in after data has been collected.

\subsection{Client wants to compute a specific value}

Clients sometimes call on a data scientist when they want to get a specific value from their data, which may be a summary statistic of one variable, or a few extreme values from a range of data points.
For instance, P1 mentioned one case where a collaborator was insistent on getting her to compute a readmission rate
in order to understand how often patients who were admitted to the hospital for a disease were later readmitted after a medical intervention:
\begin{quote}
\textit{And then there's the person who's like, ``I don't really know what [data] you need to get to calculate this readmission rate, but I just need a readmission rate. I just know that I need a readmission rate.'’ Right? You have to work with them a bit to be like, ``Okay, but what exactly do you mean by readmission?'' And you have to ask some follow-up questions.}
\end{quote}
One challenge in these situations is that the data scientist needs to establish context as to why the client wants a certain value to be computed, what data is needed for such an analysis, and whether alternative computations would be more appropriate. Again they are dropped into the middle of an analysis project without being able to design the research questions or data collection methods.

\subsection{After the client already tried to do analysis themselves}

Finally, clients may call on a data scientist after they have already tried to do part of the analysis themselves but without success, often due to their lack of data science expertise. For instance, a lab researcher in genomics may have only basic data analysis skills and unsuccessfully tried computing gene expression values for a paper; that failure motivates them to bring in someone with more experience in analyzing larger-scale data.
Surprisingly, this can cause a client to under-appreciate the amount of time and effort that data science work takes, as P10 describes here:

\begin{quote}
\textit{Sometimes people have tried it by themselves. They're frustrated. They're like, ``Oh, like this is just getting in my way. Can you just do it? And, we know you're good at this. Can you just do it fast?''}
\end{quote}

%as was the case for P10, who has worked for years as a research scientist at a leading neuroscience institute

% SKTODO: maybe some outro here
The main communication challenge here is for the data scientist to provide emotional reassurance to their client, who is likely already frustrated by their failed attempts, while at the same time setting realistic expectations that the desired work may take considerable time and effort to complete. These interactions can be sensitive since, again, clients are often in a position of power over the data scientist (e.g., it may be their research lab's PI who tried to do some analysis but failed).
In \sec{sec:magic} we further discuss how the invisibility and opaqueness of data scientists' technical work often makes clients underestimate the amount of effort involved.

\section{Problem Framing and Bridging the Gap}
\label{sec:frame}

\begin{wrapfigure}{r}{0.55\textwidth}
\includegraphics[width=0.55\textwidth]{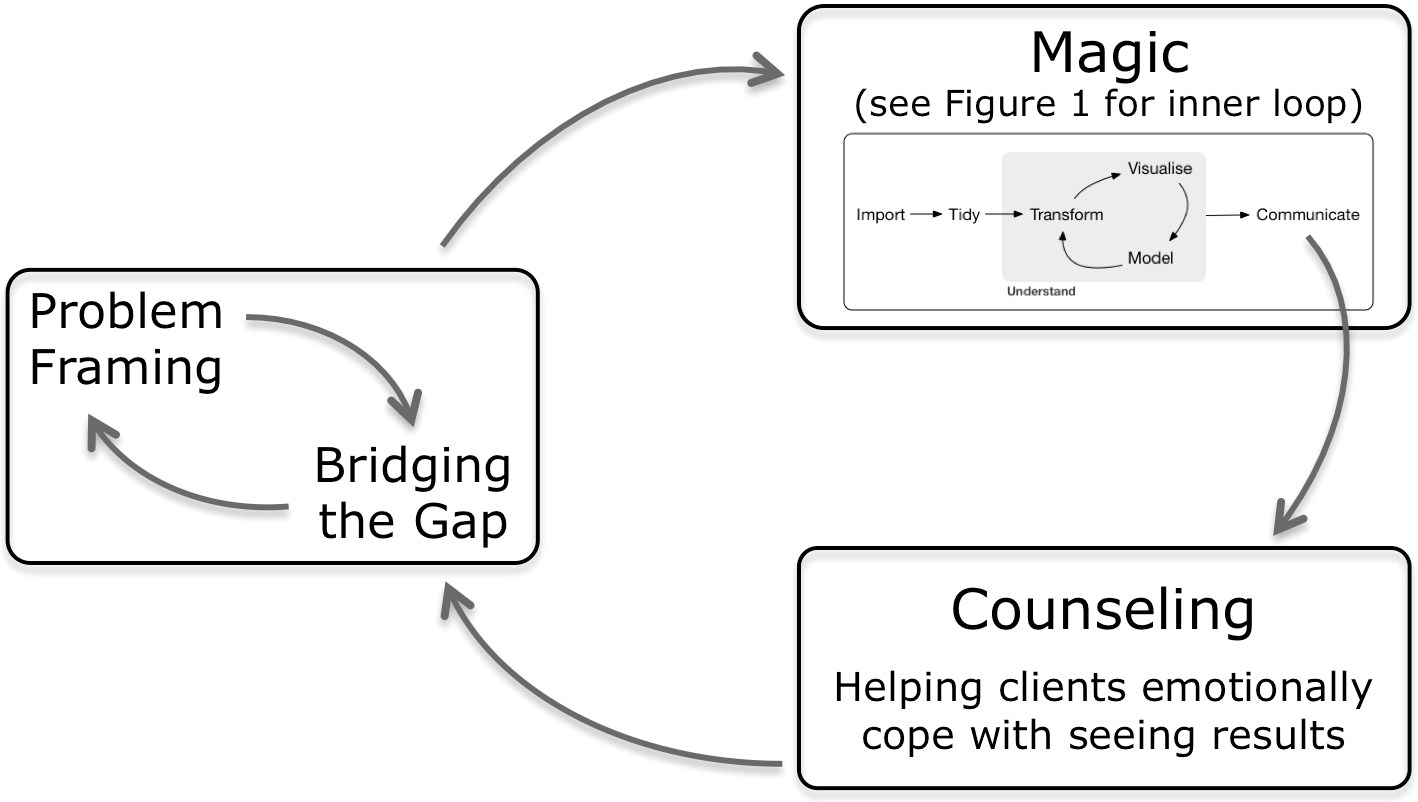}
\caption{The outer loop of collaborative data science (reproduced from \fig{fig:workflow}), with all the technical work occurring in the inner loop within the ``magic'' stage.}
\label{fig:mainloop}
\end{wrapfigure}

After orienting themselves at an entry point, the data scientist and client enter into the main loop of their collaboration. As \fig{fig:mainloop} shows, this loop involves problem framing and bridging work (this section), magic (where all the technical data analysis work occurs), and counseling the client about analysis results.

When the main loop begins, the data scientist does not start writing analysis code right away. Rather, similar to the practice of professional designers~\cite{Cross2011,Cross2018}, they first engage in \emph{problem framing} by having conversations with clients to get at the underlying problem they want to solve, which may be different than what they claim to need. During these conversations they also work together to \emph{bridge the gap} between the data scientist's analytical expertise and the client's domain expertise.

\subsection{Problem framing: asking questions to get at the underlying problem}
\label{sec:prob-framing}

%Nigel Cross designer expertise book chapter stuff:
%- "Unlike `normal' problem-solving, in a design project it is often not at all clear what `the problem' is; it may have been only loosely defined by the client, many constraints and criteria may be undefined, and everyone involved in the project may know that goals may be redefined during the project."
%- problem framing: ``Successful, experienced, and -- especially -- outstanding designers are found in various studies to be proactive in problem framing, actively imposing their view of the problem and directing the search for solution conjectures"; this happens throughout the process
%- solution conjecturing: using partial solutions as a way to define the problem better
%- co-evolving the problem definition and the solution in a cycle

All ten data scientists we interviewed engaged in
\emph{problem framing} conversations before starting analysis work, which is akin to a practice that professional designers engage in where they question the client's initial assumptions and brainstorm with them to get at their underlying issue~\cite{Cross2011,Cross2018}.
For instance, P1 explains how even a seemingly simple request for one data value can lead to a detailed back-and-forth conversation as she tries to understand what the client really wants:

\begin{quote}
\textit{I'll get a random message from someone saying, ``Hey, do you know what this number is?'' And I'll say, ``Maybe, let's say yes, I do know what that number is. But tell me why you're asking? Is there a bigger problem than I can help solve? Are you having conversations that are going to lead to more projects later, or does this reflect the fact that you don't have immediate access to this number?'' There's always a question under the question, right?}
\end{quote}

Uncovering the `question under the question' can be a tricky process since a client's ultimate end goal may not be clear at the beginning of an analysis project, and the attributes of what the data analysis should achieve may not be precisely defined. P3, who investigates digital interventions in public health, explains the kinds of questions she asks to get to the bottom of these issues:

\begin{quote}
\textit{When someone comes to me and says, ``I want to do X,'' I ask a ton of clarifying questions: What will this data be used for? Is it going to be like a one-and-done publication? Or is it going to be a platform that's rolled out like a website? Or, how far in advance do you want to predict [using some model]? Would it be successful if it was predicting one month out or are you looking for 12 months of prediction?}
\end{quote}

% PG: cut ... not super interesting:
%Clients bring a wealth of knowledge that factors into a data analysis, but P2 reported her experience asking questions as a means to wade through a client's expertise (in her case, clients are scientists who study cancer genomics) to find agreement on how to begin the analysis:

%\begin{quote}
%\textit{As a data scientist I'm very interested in kind of prodding; whenever a collaborator comes to me, I also want to prod. So their comfort zone is typically to talk about something that they're trained in, that they know very well. And they just usually lean back on that. So like for example, I had a conversation with an expert in lupus. And it was a fantastic conversation, but it took me a while to pull out the question that they were actually interested in. And they're the clinician, they see patients. It's hard to put us all on the same page, initially.}
%\end{quote}

One specific problem framing technique that some participants used was to ask their clients for what kind of output they want.
Focusing on asking questions about concrete analysis outputs like tables, graphs, and models can help data scientists productively drive conversations about what questions an analysis should try to answer, as P3 recounts from doing analysis for a public health research project:
\textit{``I asked [my client] to lay out what kind of table they would like to see, and then I'm able to move in that direction or give them something very close. And having that conversation where I say, `Tell me the ideal tables you'd like to see in your paper,' makes it so much easier for me.''}

Finally, data scientists need to be careful to respect the domain expertise of clients while communicating that their client's goals for an analysis may not be feasible. P6 outlines the difficulty of trying not to offend a client who they were working with to refine the premise of an analysis of a product A/B test:

\begin{quote}
\textit{I think the hard thing is saying [to a client]: ``I'm not questioning you or your job, I just need to understand what you're doing this for. So that I can actually do my job properly and get you the right thing rather than just giving you what you want, which probably isn't correct.''}
\end{quote}

\subsection{Bridging the gap between data science and domain knowledge}
\label{sec:bridging}

%\todo{from sean's mega-caption} In the problem framing stage, a consulting data scientist is trying to pin down exactly what questions their client is interested in answering, and how evidence to provide answers to those questions could be realized through data scientific methods. Oftentimes both the data scientist and the client must pause the problem framing stage to engage in bridging work. In this study we focused on the bridging work performed by consulting data scientists, where they try to learn more about their client's domain, and they try to explain data scientific concepts to their client. Once bridging work has been done, they can return to the problem framing stage, both equipped with a better understanding of what they are trying to accomplish.

While in the midst of problem framing, data scientists and clients often need to learn more about each others' respective expertise. We refer to this as bridging work, as it involves trying to establish common ground~\cite{Clark1991} by building bridges across the gap between data science and domain knowledge. That is why problem framing and bridging work are shown together in the same stage in \fig{fig:mainloop}: It reflects how data scientists and clients move fluidly between these two kinds of complementary activities as they try to understand both each other and the underlying problem better.

%Specifically, although the data scientist may know the statistical methods that are required to answer a scientific question, and the client can bring years of experience in their domain to an analysis, their understanding of each other's expertise may initially be low.

Bridging work occurs in two directions: the data scientist teaching the client about technical analysis methods, and the data scientist getting the client to \emph{teach them} about the target domain.

\subsubsection{Data scientists teaching their clients about data analysis methods}
The first direction is when data scientists teach clients about methodological ideas to help those clients think about how their business or scientific problem can fit into a known data analysis workflow. These conversations occur when a data scientist believes they can help their client frame their project in terms of specific statistical phenomenon or analysis methods. For instance, they may want to explain how power law distributions arise when talking to a client who is an online bookseller trying to understand sales patterns (i.e., the majority of sales will come from only the few most popular books).
Data scientists hope that clients can articulate their requests in more realistic terms if they understand the analysis tools and methods that the data scientist can feasibly use.
%
% PG: cut ...
%P1 explains how this conversation may unfold differently based on the client:
%\begin{quote}
%\textit{How I frame a question might differ depending on the situation or the person. If I know that they have some statistics training and they know what I'm trying to get at, I'll just say, ``Hey, what's the standard deviation? Do you have an estimate of that?'' [...] If they're brand new to research, and they say, ``What's a standard deviation again?'' then that's not going to work.}
%\end{quote}
%
%
P10 mentioned calibrating expectations as another benefit of this bridging work:
%P10 mentioned that another benefit of educating clients about data science methods is that it helps calibrate their expectations:

\begin{quote}
\textit{I can talk multiple languages when I talk to different people. And I think that in a lot of situations, that's also the scenario where people might not know as much about data science, right? So if they don't know the language, some of the basics, then they might have unrealistic expectations. So that's also part of my vision is setting realistic expectations.}
\end{quote}

\subsubsection{Data scientists learning domain knowledge from their clients}
The opposite bridging direction is when the data scientist wants to learn more about the client's domain to hopefully integrate the client's knowledge into the analysis and to ask more informed questions when they return to doing problem framing. Since data scientists often work in a consulting capacity with different groups of clients, each project they undertake may be in a slightly different domain and thus require specialized knowledge. P1 emphasized the importance of learning about those domains from clients so she can craft more refined analyses:

\begin{quote}
\textit{The more I learned about either the clinical domain in my [past] academic life, or now either the clinical or the business domain [in my job], it helped me ask better questions. [...] I can say, ``Oh wait, do you think this variable is maybe a confounder for this and this, because this is how they're related?'' And we can have more intelligent conversations about that, and that might inform how I build the model or how we report results.}
\end{quote}

P2 expanded on this theme by mentioning that a data scientist can unburden the client from thinking about data science issues if they can learn enough about the client's domain. That way, the client can focus their efforts on thinking about their own domain, where they are most comfortable:

\begin{quote}
\textit{If you have communicated to them, essentially, what the problems are in an understandable fashion, then in their mind they're thinking about the biology [...] They're thinking about the actual context, that domain context that they know and love very dearly, to solve whatever problem that they're after.}
\end{quote}

% 6

\section{Magic: the opaqueness of technical analysis work leads to communication breakdowns}
\label{sec:magic}

%magic is a bad thing and an indicator of a breakdown in communication. when things go wrong

After some initial problem framing and bridging work, the technical work of data analysis begins. Technical work in data science has been widely-studied~\cite{Alspaugh2019,Kandel2012,Wongsuphasawat2019goals,GuoPhD2012}, so we will not repeat those details here. In short, data scientists spend most of this time in an ``inner loop'' (e.g., \fig{fig:krebs}) writing analysis code in environments based on R, Python, or other tool ecosystems. Note that this process is iterative, so they may return to another round of problem framing after getting some analysis results and showing it to the client, as the outer loop in \fig{fig:mainloop} shows.

Prior studies of data science collaboration~\cite{Kim2016,Mao2019,Zhang2020} focused on the ways in which data science teams and other stakeholders collaborate on the technical aspects of their work (e.g., what tools they use or how data gets passed between people). Thus in this section we focus on how clients (who may not know all these technical details) perceive this work when they are not involved in the day-to-day implementation.

\subsection{Clients perceive data science work as ``magic''}

Most notably, half the participants we interviewed (P1, P3, P4, P5, P9) mentioned the word ``magic'' when describing how their clients perceived their work, without any specific prompting from the interviewer. We were surprised by the frequency with which participants talked about magic, and the way they all used the word to describe similar phenomena.
P4, who works at a research-driven medical center, took pride in his work with doctors, but also pointed out how little insight those doctors have into the data analysis process:

\begin{quote}
\textit{Usually a lot of them are clinicians, like grateful physicians, who essentially think you do magic, right? They don't know how you do it. They are never going to figure out how you do it. And they're just like, ``that's amazing'' and ``do more!''}
\end{quote}

Although P4's clients make him feel like a wizard, the opaqueness of technical work from the client's perspective can also make data scientists feel like their work is underappreciated or misunderstood. P1 expressed how she feels her analysis work is trivialized by clients since it is to opaque to them:
\textit{``There's a lot of like, you know, `You go and do your magic and then come back with the numbers.' Which I hate, I really hate that. Can you tell? I just had a meeting like this last week.''}

\subsection{How perceptions of ``magic'' lead to communication breakdowns}

Others mentioned how the `magic' aspect of data analysis work reflects a breakdown in communication between them and their client. Since most day-to-day technical analysis work is opaque to clients, it is difficult for them to understand which parts are easy and which are hard for the data scientist to complete. P3 explains how this issue manifests in her work:

\begin{quote}
\textit{We have a lot of communication issues. Those mostly center around like how requests are made to me. For example, I'll be asked for data, and it's kind of like data is this magic thing that they think I just have on my computer. Like, five minutes later they expect me to say: ``Here's your data!''}
\end{quote}

P3 continued to explain how since she had access to certain data sets that were relevant for her client, the client assumed she had equally-easy access to related data that they had not discussed before (which was not true).

% PG - cut, too roundabout and hard for people to understand ... I summarized it above
%\begin{quote}
%\textit{``Can you just look up the Google search terms for like panic and anxiety attacks or panic attacks?'' And I thought, don't react, just say yes, just tell them you can get it to them after this. But it's kind of like they don't understand that they're interrupting my workflow. And now I have to like, remember the process of how I pull that data, which is fine because I wrote it down, thankfully. Let me go pull that data for you. Oh - but in this eight minute interaction, the request changed two times. So like you come to me for a request, but you haven't really thought about that request a priori, so that I have to keep iterating.}
%\end{quote}

This opaqueness also makes it hard for clients to set realistic expectations about how long some analysis task should take. P8 works in a large federal environmental agency in North America, and here she explains the challenges of showing the client that she is making consistent progress even though much of her work is invisible to them:

% mismatch of expectations! + "my code is compiling" vibes (https://xkcd.com/303/)

\begin{quote}
\textit{When you're solving problems with data, there's this idea that you should be able to hit the ground running. And of course we all know that you could spend weeks and weeks and weeks just trying to actually get that data, and get it into a format and understand it and read metadata. And so I think one challenge of a data science team is to be able to figure ways to keep everyone engaged and show value in those early stages of a data science work or data analysis workflow that are kind of less exciting. Often not much to show.}
\end{quote}

% cut: Like, ``Hey, I've got a big clean tabular data frame and that took us four weeks to clean'' and your stakeholders are like, ``Really? I thought we already had the data!''

The tremendous effort involved in acquiring and cleaning data has been well-studied, and it is widely known that it amounts to a significant portion of a data scientist's working time~\cite{Kandel2011,Kandel2012}. However, this fact is still not visible to many clients, as P1 explained when telling the interviewer about a time she spent two weeks cleaning and wrangling human-generated free-text data:

\begin{quote}
\textit{If you ask our business stakeholders what I've been doing this whole time, and they would say, ``I actually don't know. I think she's working hard. She seems busy, but I don't really know what she's doing.'' That work is not visible to them, nor should it be. This is not for them to consume. This is to build the foundations of a product that will then be for them to consume. But it's very necessary.}
\end{quote}

In sum, since the day-to-day technical work seems ``magical'' and opaque to many clients, they often have a hard time communicating productively with data scientists until they can see some partial results (see next section), which may not come for a few weeks at a time.

\section{Counseling: Showing results to clients and helping them cope}
\label{sec:counseling}

%Communicating results to client and managing their emotional responses}

%Coping: managing the client's emotional reactions to presented results

After some period of technical analysis work (e.g., a few days or weeks), the data scientist shows some results to their client. This section describes the most salient types of communication that occur when the client sees these partial results.

\subsection{Helping clients cope with unexpected analysis results}

Without prompting, many of our participants mentioned doing some informal counseling during result presentation meetings to help clients ``cope'' with unfavorable results. Like in many fields of science, technology, and design, there may be no data or experimental specification that will yield a result that clients are happy to receive. In data science, clients may want to show via a data analysis that their vaccine candidate is effective at preventing disease, that a governmental policy has reduced air pollution, or that they can predict with high accuracy which customers will spend the most money on their e-commerce site. But it is not clear whether these aims have been achieved until a data scientist has completed their analysis after a span of weeks or months.
%
%showing a data-driven result requires confronting fundamental truths about nature and human behavior which no amount of data or sophistication of analysis can overcome.
%
Even though P7 has over 20 years of experience working as a quantitative geneticist, he still struggles when delivering results that clients will not like to hear:
\textit{``How to present bad news? Um, I don't really know how to do it. And I would say from my experiences in my career, I still haven't figured it out.''}

Data scientists help clients cope with unfortunate circumstances throughout the course of an analysis, not just when presenting results. For instance, P2 was brought in after the data had already been collected, but then she had to tell her client about a serious flaw she eventually discovered:

\begin{quote}
\textit{Some respond better than others. I've had one extreme with a collaborator who we discovered had a confounded study design. They had already done the experiment and so we had to go to them and explain to them what happened. They were all excited about the biology, but once we told them, they went through the full seven stages of grief in that one hour.}
\end{quote}

Also, data scientists acquire their training from a variety of academic fields~\cite{Kross2019}, which often have different norms for canonical statistical methods, reporting of findings, and thresholds for significance of results. Here P9 recounted his experience doing counseling (`playing therapy') to reassure a client of the soundness of his results when the client had  different field-specific norms:

\begin{quote}
\textit{There's definitely also like a tension of field-specific expectations or norms and your principles or values as an outside consultant or analyst. And then it gets manifested as like ``my analyst did something wrong'' or ``they didn't do what I told them.'' And then you just sort of feel like you're playing therapy with them, trying to reason out why what you did is actually okay.}
\end{quote}

\subsection{Power dynamics when presenting results to clients}

Power dynamics appear here as well: Some of our participants work in environments (e.g., large companies) where there are many data scientists that clients can consult with. Therefore, they are afraid that they may be `replaced' if they deliver results that a client does not like. P3 laments that clients do not understand how much they may be giving up by finding a new data scientist (whom they hope will produce more favorable analysis results) rather than sticking with the current one:

\begin{quote}
\textit{When someone doesn't get the result they want as fast as they want, they think that it could be your fault. And so they say, ``Well let's just bring in another one of you!'' But they don't understand that I can't just give you my code. They don't understand that it doesn't just transfer without any effort, like the other person has to understand what's going on.}
\end{quote}

Ultimately, helping clients cope with analysis results comes down to the data scientist's ability to empathize. They often have to find a balance where they are not so empathetic as to make undesirable results look better than they actually are, but to still be mindful about how clients are going to be affected by bad news.
For instance, P1 explained her approach to empathy for her clients and how she frames analysis results in terms of how it will affect her client's future plans:

\begin{quote}
\textit{If something doesn't go the way that somebody likes, how do you handle that? I think I'm a pretty empathetic person. And so being able to put myself in their shoes and think, ``Ugh, this might mean they're not going to get that grant.'' Or, ``Oh, this means the hard work that their team put in is not reflected in our data. And therefore we're not going to be able to prove that [a medical intervention] helped.''}
\end{quote}

In sum, the main loop in \fig{fig:mainloop} continues until the end of the client engagement. Ideally there is an outcome that satisfies the client, but sometimes data scientists must help clients cope with a less-than-desirable ending. Here P8 reflects on different ways that engagements can end:

\begin{quote}
\emph{Success is not just the thing you build, but was it a successful process? And many times things are not successful in that the analysis either didn't get finished or you didn't find anything, right? One of the things about looking for patterns in data is like, well, we look for patterns in data and there were no patterns.}
\end{quote}

\section{Participants' thoughts about data science education}
\label{sec:dse}

For our final set of findings, we report on an unexpected theme that many participants raised. Even though the focus of our study was not on data science education, we found it interesting that the majority of our participants (P1, P2, P3, P4, P6, P8, P10) brought up this topic during their interviews without prompting. Specifically, they recalled that they either spent no time or minimal time learning how to navigate the socio-emotional work of data science (i.e., the `outer-loop' workflow stages we have described so far) during their formal education. For instance, P1 said:

\begin{quote}
\emph{I think a lot of it, you can't really learn in school. It's hard to not learn it on the job. [...] There's so much that I could not have learned in grad school about how to work with stakeholders, the psychology of it all, like when to use what kind of conversational style or rapport or how to report my results.}
\end{quote}

P2 mentioned how the teaching of software tools for doing data science is well-understood in academia, but that is not sufficient for helping students to become effective data scientists:

\begin{quote}
\emph{We are master tool teachers. We teach tools, we teach software, we teach methods. But by no means does that mean when a student walks out the door, they know what it means to produce a good data analysis. [...] We have to be able to tell students in a 12 to 16 week course what it means to be able to have that initial conversation [with a client] in an effective manner, and then how to get them to build a `successful' data analysis and an effective, efficient manner.
} \end{quote}

In \sec{sec:dse-discussion} we discuss some implications of these findings for data science education.
\section{Discussion}
\label{sec:discussion}

We reflect on our study findings in terms of its relationship to prior work in CSCW, the socio-emotional infrastructure that our clients must build, implications for education, the parallels between data science and design work, and implications for tool design.

\subsection{Relationship to Prior CSCW Work}
\label{sec:big-one}

Our findings enhance the perspective of data scientists' involvement in computer-supported cooperative work. In the following subsections we discuss how our findings in all stages of our outer-loop workflow from \fig{fig:workflow} both complement and extend relevant prior work in CSCW.

\subsubsection{Groundwork}

During the Groundwork stage, data scientists try to build relationships and establish trust with clients before a collaboration can even begin. Sometimes these relationships are already arranged beforehand; for example, the participants in the studies of Mao et al.~\cite{Mao2019} and Hou and Wang~\cite{Hou2017} already have relationships with clients within the context of an ongoing research consortium or a hackathon, respectively, so those studies did not cover the Groundwork or Orienting stages (next subsection) that we did. Mao et al.\ contributed novel insights about building common ground during data science collaborations~\cite{Mao2019}, and we extend that notion to laying groundwork before these collaborations even begin.

The importance of establishing trust is also explored in work by Lawrence~\cite{Lawrence2006} and (separately) Tang~\cite{Tang2019}, who point out that trust in the beginning of client-consultant relationships is dependent on reports from third-parties and overall reputation. Our work extends their findings by adding firsthand descriptions of how data scientists do their own ``marketing'' to build their reputations both within their organizations and externally in the data science community. Lawrence showed how large e-Science research projects are facilitated by years-long relationships and foundations of trust that started long ago when leaders of those projects were training together or had advisor-advisee relationships in formal Ph.D. programs~\cite{Lawrence2006}. Our work extends those findings by showing what needs to be involved at the very beginning of client-based relationships. As data science work becomes more widespread and democratized, fewer people will be able to rely on traditional relationships fostered during long-term academic training or Ph.D.-level professional networks; therefore, understanding how these cooperative relationships start in diverse settings is important for improving data science practices.

\subsubsection{Orienting}
This is the process through which data scientists first enter into collaborations and `find their bearings.' Since our participants are well-established, with 5 to 25 years of experience (\sec{sec:methods}), their collaborations start when a client approaches them through one of the five entry points from \sec{sec:orienting}. However this is not always the arrangement in all data science collaborations. For example, Choi and Tausczik describe the experiences of citizen scientists, journalists, and volunteers working on open civic data projects that either directly benefit a community or their local government~\cite{Choi2017}. In their case the people working as data scientists are the ones initiating the collaborations, and often their progress on a project is self-directed rather than client-directed. Mao et al.\ describe a few large-scale crowdsourcing collaborations where many domain experts are consulting with many data scientists; those data scientists reported that it could be difficult to find entry points for projects because many collaborations are short-lived, which does not give them enough time to develop relationships that would facilitate follow-up contact with previous clients~\cite{Mao2019}. Overall, our work extends prior CSCW studies by characterizing the orienting activities of data scientists who have worked in years-long collaborations through many project cycles.

\subsubsection{Problem Framing}
The process of Problem Framing involves asking the right questions: Asking questions allows data scientists to narrow down exactly what data-informed question a client is interested in addressing, which then allows data scientists to propose technical methods that could be used to answer this distilled question. Lawrence's findings show that it can be unclear when to ask questions and whose role it is to ask and answer questions when large groups of people are collaborating in multidisciplinary e-Science projects~\cite{Lawrence2006}; in our setting the roles are more clear-cut since the data scientist takes the lead on problem framing. Mao et al.\ focus extensively on question-asking: among their most important findings is that asking the right question can lead to new lines of research and scientific discoveries~\cite{Mao2019}. We extend those findings by showing that what separates Problem Framing from just asking the right questions is that Problem Framing is a methodical process that encompasses more than honing in on questions: it also helps them plan for future products and find agreement with clients about deliverables (e.g., what graphs or tables do they want to see?); we also connected this process to how professional designers work~\cite{Cross2011,Cross2018}. In addition, our findings in this stage extend prior CSCW study findings by revealing the emotional labor~\cite{Grandey2000} that data scientists have to engage in to make their clients feel at ease when put in the uncomfortable position of frequently questioning clients who are often in positions of power.

\subsubsection{Bridging the Gap}
To bridge the gap between data and domain expertise, Mao et al. reported that data scientists in their study frequently asked their collaborators (biomedical domain scientists) about what the ambiguous variables in their data represented~\cite{Mao2019}. Third parties can also serve as bridges, like how Hou and Wang's ``client teams'' acted as brokers during a data hackathon for nonprofit organizations to inform volunteer data scientists about the origin and domain of their data~\cite{Hou2017}. Additionally, Choi and Tausczik describe open civic data projects where domain experts explain the features of government data to data scientists~\cite{Choi2017}. In all of this prior work, though, most of the knowledge transfer \emph{is in one direction}: from domain expert to data scientist. We extend these prior findings by showing how the data scientists in our study revealed that they provide the social and contextual scaffolding necessary for teaching and learning in both directions: Our participants reported that this both helps them understand the client's domain (which has been well-covered in prior work) and also transfers knowledge from data scientist to domain expert. Most notably, we discovered that data scientists can help their clients to work more effectively with them if they proactively educate clients about the capabilities and limitations of relevant data-analytic methods.

\subsubsection{Magic}

The problem of opaqueness pervades many kinds of data science collaborations. For instance, Mao et al.~\cite{Mao2019} and Hill et al.~\cite{hill-2016} describe the pervasive interpretability problems of ``black box'' methods, where project collaborators, or in some cases the data scientists themselves, are unable to understand or explain the results produced by data science or machine learning algorithms. While this is an important problem, it is different from what our participants meant when they mentioned `magic' during interviews. Our participants meant that clients perceived an opaqueness \emph{in the actual process of how data analysis work was done}, even when the outputs were interpretable. Thus, instead of focusing on the opaqueness of algorithms (which is well-covered by prior work), our findings shed new light on the opaqueness that clients perceive in the data analysis process. This sometimes results in their dismissive attitudes about technical work (e.g., quotes like ``go do your magic'') as clients chose to exclude themselves from the nitty-gritty of technical processes. This attitude frustrated our data scientist participants since it belittled their role in the project, as if they were themselves a ``black box'' like an algorithm. It also hindered communication, since technical decisions during data analysis project may need finer-grained input from clients. Lawrence touches on some of these issues in her findings about the complexities of communication in a distributed e-Science consortium~\cite{Lawrence2006}. In sum, clients' \emph{perceptions about} and \emph{willingness to participate in} technical data analysis work can dramatically affect both the work itself and the emotional state of data scientists who work with them.

% PG cut out: That said, based on reports from our participants we believe that clients' past experiences with ``black box'' algorithms may cast a shadow of impenetrability over the entire technical data science process.

%\todo{Sean can you say a bit about interpretability in ML that R2 mentioned? make distinctions between opaqueness of ML model and opaqueness of data science process. 'TODO: Actually this is a nice opportunity to draw the distinction between opaque processes and opaque products.' from R2: 'the black-box nature of data science is also well-known, which has popularized the area of ``interpretability’’, and there is a large body of model exploration tool concerns about communicating the model results to clients.'}

\subsubsection{Counseling}
Prior studies of data science collaborations~\cite{Mao2019,Hou2017,Zhang2020,Kim2016} have not yet investigated the emotional labor~\cite{Grandey2000} required when data scientists present unexpected or sub-optimal results to clients. Thus, our findings add to this literature by highlighting what goes on in the Counseling stage. Our investigation of these issues is influenced by Slov\'{a}k and Fitzpatrick's call for more integration between CSCW and lessons from the field of social and emotional skills development~\cite{Slovak2015}, and Hochschild's work on the role of emotional labor as a professional skill~\cite{Hochschild2012}. Our findings also include participants' reports about their own feelings as the bearers of bad news, and how their awareness of the attention and resources that their clients have put into a project affects how they frame potential disappointments for clients. This contribution is novel within the literature at the intersection of CSCW and data science, but more broadly it follows prior work by Moncur about how HCI researchers protect themselves from emotional harm~\cite{Moncur2013} and work by Raval and Dourish about the emotional labor that rideshare drivers perform for their clients~\cite{Raval2016}.

\subsection{Socio-Emotional Infrastructure for Data Science}

The outer-loop workflow that our findings reveal (\fig{fig:workflow}) consists not of technical activities but rather of \emph{socio-emotional} activities such as developing trust, building common ground, and providing emotional support to clients throughout the lifetime of a project.
In addition to all of the technical work required to be a professional data scientist, our participants conveyed how they navigate relationships with their clients and the work culture in which they are situated, as well as how they manage emotions and reactions that arise during an analysis. Our investigation shows that relationship-building and maintenance with clients is made up of many systematized interactions that are common and repeated both during an individual data scientist's career and across data scientists working in different domains in industry and academia. We characterize this kind of work as \emph{socio-emotional infrastructure} --  interpersonal scaffolding that data scientists are expected to provide for clients. This sort of human-to-human infrastructure complements the technical infrastructure that has been widely-studied in prior work, such as computational notebook platforms~\cite{LauVLHCC2020}, cloud computing services, and I.T. infrastructure for data processing ~\cite{DSDE}.

According to our participants, a major purpose for providing this infrastructure is to avoid dysfunction and encourage better practices throughout an analysis. For example, in the orienting stage a client may have an analysis method they want to use that the data scientist believes is inappropriate given the analysis question they are trying to answer. In this case, the data scientist may need to spend some of the social capital they accrued earlier (during the groundwork stage) to help the client avoid the mistake of using a method that may be too costly or technically burdensome. In this way the data scientist is lending their expertise as infrastructural support for making decisions about an analysis, which can lead to better results and cost savings down the line.

Without this analytical expertise and contextual support, clients would otherwise be left to navigate these decisions themselves. Considering the democratization of data science tools and the proliferation of AutoML~\cite{WangDakuo2019} (i.e., AI support for data science) and `push-button' cloud-based data science products, it is easier than ever for an end user to upload their data to one of these services and get results without ever interacting with a human being. Although this method of doing data science may have less friction and lower costs, there is little guidance in these products to indicate whether the end user is using appropriate methods for their circumstances, whether the methods are working as they intended, whether the user has the right data to answer the questions they are interested in, or how their results should or should \emph{not} be interpreted. We believe our findings show how \textbf{\emph{the socio-emotional infrastructure that data scientists provide will not be easily automated away}}, even as increasingly sophisticated AutoML methods become commoditized.

\subsection{Implications for Data Science Education}
\label{sec:dse-discussion}

Even though the focus of our study was not on data science education, many participants brought up this topic, which we describe in \sec{sec:dse}. Most notably, they mentioned a lack of training in socio-emotional skills that surround the technical work. Considering how these socio-emotional skills are vital to our participants on the job, we believe that data science curricula should place more emphasis on teaching them. This is a known gap in current curricula, since a recent survey of data science course descriptions does not mention any focus on building these skills~\cite{kross-ddse}. We hypothesize that socio-emotional skills may be rarely taught in data science programs because the concepts required to excel in the outer-loop workflow of \fig{fig:workflow} are more challenging to teach than, say, specific programming languages, technical tools, or statistical methods. But tools-related knowledge is increasingly becoming commoditized via free online tutorials, MOOCs, and data science coding bootcamps. However, what is much harder to commoditize are the subtle interpersonal dynamics that occur between data scientists and their clients, so perhaps where formal education can add the most value is in fostering these skills in a more rigorous way.

%A curriculum of programming projects and statistics exams, is not intuitively congruent with teaching socio-emotional skills. Passing unit tests and grading multiple choice questions are much more straightforward evaluations of students' knowledge compared to understanding how they interact with clients.`

\subsection{Data Scientists as Designers}
\label{subsec:design}

Our study findings highlight the ways in which data scientists function as designers and co-designers of analyses when working alongside their clients. We noticed that many of the conversations and behaviors they described run parallel to the kinds of design processes that have been studied by ethnographers of design~\cite{Cross2011,Cross2018}; some data science practitioners have also discussed these parallels in research papers, blog posts, and podcasts~\cite{NSSD,Peng2019a,Peng2019b, Stoudt2020principles}. For example, our participants reported needing to reconcile many competing interests -- including the client's ambitions for a project, the limitations of available data sources, and constraints about the time and resources available to work on an analysis. Ultimately data scientists need to strike the right balance between all of these factors to create a coherent analysis that meets their own standards for ethics and rigor while also properly serving their client's needs. Designers must similarly manage this socio-emotional process of negotiating with stakeholders and working within prescribed limitations on client projects.

Like designers, data scientists help their clients approach problems via the iterative process of problem framing~\cite{Cross2011}. Several of our participants mentioned how even when a client would approach them with a problem that the client thought was well-defined, they would still begin with a series of questions to get the client to think more deeply about their prior assumptions. This made clients realize that their initial problem was unclear, and then our participants would work with them to re-frame their problem and revise their goals. Similar to how designers work, this style of breaking down problems and then re-framing them means that data scientists are often embracing uncertainty and helping clients work through the resulting ambiguity. Coping with uncertainty is such an important skill in data science that it has been incorporated into some data science courses~\cite{Kross2019}. Also like designers, data scientists need to cyclically return to the problem framing stage if new data or new constraints are introduced as an analysis proceeds (\fig{fig:mainloop}).

In addition, both data scientists and designers draw upon a repertoire of examples from past projects that they creatively adapt to new challenges. For a data scientist, starting a new analysis project by seeing how it is similar to a past project of theirs can help constrain the design space of techniques to apply. However, a problem faced by data scientists that may not be as common in other design disciples is the lack of publicly-available analysis examples due to sensitive or proprietary data~\cite{Peng1226,hicks2019elements}. Thus, there are missed opportunities for data scientists to learn from the experiences of \emph{others}, whereas designers in many fields can draw upon many publicly-visible examples to learn from (e.g., in visual design, user interface design, industrial/product design).

Finally, like designers, data scientists use various concrete representations to aid in problem framing, bridging knowledge gaps, and discussing results with clients. In design studies, visual representations are often used for communication, namely for creating a shared space to work out ideas and make them more concrete~\cite{Cross2018}. Data scientists frequently create graphs, diagrams, tables, statistical models, and computational notebooks as representations to communicate with clients. In the next subsection we will discuss some ways that these current representations fall short.

%- data scientists do not get the training they require to manage the socioemotional workflow
%- so much curriciulum is geared towards tools and algorithms because you can teach it, this outer loop stuff is hard to teach so people don't teach it
%- how do we get to the point where we could teach a cogs127-lime course exceot about data science

\subsection{Tool Design: Lack of Expressive Representations for Collaborative Data Science}
\label{subsec:represent}

%\todo{flip the script here about how most study papers end with implications for tools ... but we're more 'pessimistic' right now?}

%Add deepnote comparison table 

%We are not suggesting that you should build one tool to try to solve this problem. Re-branding figma will not work.

%Prior papers have suggested all of these `inner loop tools' like slack, asana, jupyter, etc. Those help with the inner loop but not the outer loop.

We conclude by discussing some implications of our study findings on tool design for collaborative data science.
In the prior section we discussed ways in which data scientists work like  designers. One salient attribute of design work is the ability to use representations as an external memory for the working state of a problem and as an artifact to facilitate communication with clients. For instance, in visual design a common representation may be an Adobe Illustrator or Photoshop file that resides in a shared folder. \emph{The communication challenges in our study findings suggest that that data scientists may lack representations with the same expressiveness in facilitating client interactions as those used in other design fields.}
To show a concrete example of this contrast, we will compare the work of data scientists to that of product designers and web designers.

Product designers use computer-aided design (CAD) software to create 3D models of an object that will eventually exist in the real world. They use these models as prototypes to see if the object fits a client's specifications, and clients can easily see and comment on parts of the object's design that may need to be changed. Product designers can also place their models into simulated environments so that clients can see how the object would look in its intended environment (e.g., on a kitchen counter); they can even 3D-print low-fidelity prototypes to place in real physical environments.
%
%Simulated environments and their real-world constraints can directly inform decisions made by the designer and their client.
%
%An extreme but illustrative example is the case of an antennae design for a spacecraft, where the electromagnetic properties of a signal and the constraints of the project informed computer models that generated visually un-intuitive but functionally-ideal designs. \todo{CITE}
% CITE: https://link.springer.com/chapter/10.1007%2F0-387-23254-0_18?LI=true
%As an extreme example, aerospace engineers can test how an aircraft component acts in a simulated environment, and they can even use properties of that environment to dictate the design of a component.
%
%The product designer (and the computational fluid dynamics engineer, in the case of the satellite antennae) are using computers to model a real-world object that they and their clients have common reference point for in the physical world; thus, they can  encompassing its use, abilities, and limitations.
%
Similarly, web designers prototype their work using computer-aided `CAD-like' tools such as Figma, Sketch, or Adobe XD. Even though these designs exist purely as pixels under glass in a virtual world, there is still a known set of physical analogues such as space layout constraints, UI widgets such as scrollbars and menus, and interaction physics like scroll momentum and touchscreen gesture responsiveness. With tools like Figma, web designers can make prototypes look and feel `real enough' for clients to give rich feedback on it throughout the design process.

Tools like CAD and Figma both allow designers to work at the level of their expert understanding, while still allowing clients to intuitively experience the artifact that is being designed (e.g., a 3D object or a responsive mobile website). For example, in Figma a designer can manually specify advanced CSS rules, yet clients can intuitively grasp the effects of CSS changes on-screen. This allows clients to easily give feedback like \emph{``Can you make this menu bar a darker shade of blue and a bit taller?''} without needing to understand the technical details of how CSS layouts work.

In contrast, data scientists still lack tools like CAD, Photoshop, or Figma which would make their ongoing work so transparent that their clients can ask the data science analogues of \emph{``Can you make this menu bar a darker shade of blue and a bit taller?''}
Although hundreds of tools, frameworks, libraries, and computational notebook technologies exist for data scientists, these representations are made to facilitate the productivity of those who are analyzing data, \emph{not} those who need to consume the eventual analysis results. For instance, if a data scientist walked into a client meeting, opened up a Jupyter notebook filled with hundreds of lines of specialized Python code coupled with a CSV file with thousands of rows, then that would probably not be a very productive meeting! Instead, data scientists must painstakingly distill these opaque representations into summary charts, graphs, and tables before showing clients, which ends up losing a lot of richness of the original `uncompressed' representations. In contrast, designers can show the original CAD, Photoshop, or Figma files to clients and engage in meaningful conversations about those \emph{primary} representations.

%data scientists do not have specialized tools for the socio-emotional parts of their work.

Data scientists also need to do bridging work with clients (\sec{sec:bridging}) without the benefit of rich representations. This work requires the data scientist to educate the client about how certain methods are used in an analysis, or how statistical phenomena affect the client's problem. Our participants reported significant difficulty in explaining these concepts to clients, often relying on their own experiences teaching and learning statistics to convey a concept. They could not benefit from basic design representations like sketches or drawings, because clients often do not have enough of an understanding of data science methods to understand the statistical concepts that underlie these drawings (e.g., properties of statistical distributions or how certain factors interact).

%Similarly, our participants lacked representations when discussing results with clients and counseling them about how to move forward. Model results that report odds ratios and confidence intervals are not easily interpreted by clients, and our participants often had to deliver disappointing news about results obfuscated by statistical complexity. In other words, our participants had no representation that accurately conveyed results that could be intuitively understood by their client.

%Finally, the opaqueness (``magic'') of the data analysis process caused communication breakdowns between our participants and their clients, which might have been reduced or avoided altogether if our participants had better representations of their work-in-progress. There are many small details of creating a data analysis that are not relevant to clients, but there is an open question of where the line should be drawn in a representation of a data scientist's work so that clients have more insight into how they are making decisions during a data analysis.

In sum, we believe that in order to create collaborative tools to facilitate the outer loop stages discovered by our study, we need to invent more expressive representations.
Existing CSCW tools for data science such as real-time collaborative Jupyter notebooks~\cite{LauVLHCC2020}, version control, and project management systems help coordinate the inner-loop of technical work, but those representations are still at the code and data level, which is not as useful for interfacing with clients.
Unlike in product or web design, in data science there is often no shared point of reference between the data scientist and their client: Although data are collected from observations in the physical and digital world, the ongoing work in analyzing that data often occupies an abstract mathematical world that remains opaque (and thus ``magical'') to clients. It is an open question how we might design a tool with the transparency of CAD, Photoshop, or Figma for the outer loop of collaborative data science; our  hunch is that it will require a representation that is richer than code or data.

%We are not suggesting that you should build one tool to try to solve this problem. Re-branding figma will not work.

%Prior papers have suggested all of these `inner loop tools' like slack, asana, jupyter, etc. Those help with the inner loop but not the outer loop.

%Where a web application designer could prototype a new feature with specialized software, data scientists

%show how a new feature might be integrated into an existing product, data scientists have to work in their 

% 2AC: This is perhaps the biggest revision I'd like to see: the authors conclude with the "implication for design" that data scientists don't have tools like Figma, etc, to use with their work. Yet, they started the paper off talking about one common visualization tool that is used in many fields: the workflow diagram. There has even been prior work in CSCW and adjacent fields talking about the use of workflow diagrams as boundary objects between different stakeholders, and as important ways of recording data provenance. The authors should review this and related literature and revise this section accordingly. I'd also like to know if the participants reflected on their own use of these inner loop workflows in communicating with clients. Did any of them discuss this? or talk about specific tools they'd like to see developed?
%
% Paper of potential interest:
%
% Dourish, P. (2001). Process descriptions as organisational accounting devices: the dual use of workflow technologies. In Proceedings of the 2001 International ACM SIGGROUP Conference on Supporting Group Work, 52–60. 10.1145/500286.500297

Perhaps one step toward these richer representations is to reify workflow diagrams such as ours in \fig{fig:workflow} and finer-grained versions of it into a workflow system~\cite{Dourish2001}. Since the early days of data science, workflow diagrams have been used to track data provenance and manage collaborative work in a data-centric way, such as in graphical workflow programming systems like Kepler~\cite{Kepler06}, Taverna~\cite{Taverna06}, and VisTrails~\cite{VisTrails08}. We could potentially generalize this technology to track progress in the `outer-loop' stages that we identified in this paper in addition to the `inner-loop' mechanics of data flows. Doing so would turn workflow diagrams into boundary objects akin to design sketches or UX flows in user-centered design collaborations~\cite{Blomkvist2015}. Our envisioned diagrams could be embedded into contemporary workflow systems such as Asana, Notion, or Jira\footnote{\url{https://asana.com/}, \url{https://www.notion.so/}, \url{https://www.atlassian.com/software/jira/}} so that data scientists can keep their clients up-to-date with the state of their project; this could also help to reveal some of the opaque `magic' throughout the analysis process. However, as Dourish, Suchman, and others in the CSCW community have pointed out, workflow technologies risk hindering flexibility in creative work (such as those performed by the data scientists that we studied) because they tend to categorize work into fixed classes of tasks~\cite{Dourish2001,Suchman1993}; moreover, usually those with more power are the ones determining the categories~\cite{Suchman1993}. Thus, we must be careful to design a workflow technology that preserves such flexibility while still providing sufficient process transparency for clients.

\section{Conclusion}
\label{sec:conclusion}

This paper has argued via an interview study that we should view collaborative data science work beyond the usual lens of technical workflows like those in \fig{fig:krebs} that now pervade the field. Rather, we zoom out to characterize the interpersonal dynamics between data scientists and their clients by synthesizing a novel outer-loop workflow to capture this form of collaborative work. This workflow involves laying groundwork by building trust and reputation before a project begins, orienting to five possible entry points of client engagements, problem framing, bridging the gap between data science and domain expertise, technical magic, and counseling to help clients emotionally cope with analysis results. 
\rev{
A unique aspect of our study is that
interview participants working in both industry and academia provided their insights, thus moving the conversation forward about the professional needs of data scientists beyond the usual suggested refinements for technical tools. We believe that their perspectives illustrate a broad spectrum of experiences that can inform the intersection of data science and design practice. One potential avenue for future work is to dig deeper into possible differences between industry and academic outer-loop workflows. By studying practitioners in both settings, we aim to identify common problems (and notable differences) so that we can propose generally-applicable solutions to challenges in outer-loop workflows.}
We conclude by showing how this workflow is not well-supported by current data science education, tool development, \rev{or theories of data science practice, and we look forward to addressing those challenges in} future work.

\newpage % stent

%% The acknowledgments section is defined using the "acks" environment
%% (and NOT an unnumbered section). This ensures the proper
%% identification of the section in the article metadata, and the
%% consistent spelling of the heading.
%\begin{acks}
%\todo{comment this section out for anonymized submission}
%\todo{put names of people you've showed the draft to or discussed this work with here so you don't forget to list them later}
%\todo{thank all funding agencies and sources}
%\end{acks}

\begin{acks}
This material is based upon work supported by the National Science Foundation under Grant No. NSF IIS-1845900.
\end{acks}

%% The next two lines define the bibliography style to be used, and
%% the bibliography file.
\bibliographystyle{ACM-Reference-Format}
\bibliography{references}

\end{document}